\documentclass[%
 reprint,
 amsmath,amssymb,
 aps,
 superscriptaddress
]{revtex4-2}

\usepackage{graphicx}
\usepackage{dcolumn}
\usepackage{bm}
\usepackage{xcolor}

\usepackage{url}

\usepackage{multirow} 
\usepackage{tabularx}
\newcolumntype{Y}{>{\centering\arraybackslash}X}

\usepackage{setspace}
\usepackage{soul}
\DeclareUnicodeCharacter{2212}{-} 

\begin{document}

\title{
On the effect of electronic stopping in molecular dynamics simulations of collision cascades in Gallium Arsenide
}

\author{Johannes L. Teunissen}
\affiliation{Univ. Lille, CNRS UMR 8520 - IEMN - Institute of Electronics, Microelectronics and Nanotechnology, F-59000 Lille, France}
\affiliation{Royal Belgian Institute for Space Aeronomy, Av Circulaire 3, 1180 Brussels, Belgium}
\email{jlteunissen@gmail.com}

\author{Thomas Jarrin}
\affiliation{CEA, DAM, DIF, F-91297, Arpajon, France}
\author{Nicolas Richard}
\affiliation{CEA, DAM, DIF, F-91297, Arpajon, France}
\affiliation{Universit\'e Paris-Saclay, CEA, Laboratoire Mati\`ere en Conditions Extr\^emes, F-91680 Bruy\`eres-le-Ch\^atel, France}

\author{Natalia E. Koval}
\affiliation{CIC Nanogune BRTA, 20018 Donostia-San Sebasti\'an, Spain}

\author{Daniel Mu\~noz Santiburcio}
\affiliation{CIC Nanogune BRTA, 20018 Donostia-San Sebasti\'an, Spain}
\affiliation{Instituto de Fusi\'on Nuclear ``Guillermo Velarde'', Universidad Polit\'ecnica de Madrid, c/Jos\'e Gutierrez Abascal 2, 28006 Madrid, Spain}

\author{Jorge Kohanoff}
\affiliation{Instituto de Fusi\'on Nuclear ``Guillermo Velarde'', Universidad Polit\'ecnica de Madrid, c/Jos\'e Gutierrez Abascal 2, 28006 Madrid, Spain}
\affiliation{Atomistic Simulation Centre, Queen's University Belfast, Belfast BT71NN, Northern Ireland, United Kingdom}

\author{Emilio Artacho}
\affiliation{CIC Nanogune BRTA, 20018 Donostia-San Sebasti\'an, Spain}
\affiliation{Donostia International Physics Center DIPC, 20018 Donostia-San Sebastián, Spain}
\affiliation{Theory of Condensed Matter, Cavendish Laboratory, University of Cambridge, Cambridge CB3 0HE, United Kingdom}
\affiliation{Ikerbasque, Basque Foundation for Science, 48011 Bilbao, Spain}

\author{Fabrizio Cleri}
\affiliation{Univ. Lille, CNRS UMR 8520 - IEMN - Institute of Electronics, Microelectronics and Nanotechnology, F-59000 Lille, France}

\author{Fabiana Da Pieve}
\thanks{currently working at the European Research Council Executive Agency (ERCEA)}
\affiliation{Royal Belgian Institute for Space Aeronomy, Av Circulaire 3, 1180 Brussels, Belgium}

\singlespacing

\date{\today}

\begin{abstract}
   Understanding the generation and evolution of defects induced in matter by ion irradiation is of fundamental importance to estimate the degradation of functional properties of materials.
   Computational approaches used in different communities, from space radiation effects to nuclear energy experiments, are based on a number of approximations that, among others, traditionally neglect the coupling between electronic and ionic degrees of freedom in the description of displacements. In this work, we study collision cascades in GaAs, including the electronic stopping power for self-projectiles in different directions obtained via real time Time Dependent Density Functional Theory in Molecular Dynamics simulations of collision cascades, using the recent electron-phonon model and the previously developed two-temperature model.
We show that the former can be well applied to describe the effects of electronic stopping in molecular dynamics simulations of collision cascades in a multielement semiconductor and that the number of defects is considerably affected by electronic stopping effects. The results are also discussed in the wider context of  the commonly used non-ionizing energy loss model to estimate degradation of materials by cumulative displacements.
\end{abstract}

\maketitle

\section{Introduction}

The description of radiation-induced effects in materials is a challenging multiscale problem at both time and length scales. 
At very short time scales ($\sim$100 as), a particle passing through a material loses
energy by transferring it to electronic degrees of freedom of the target (electronic stopping).

On longer time scales, predominantly when the particle has been slowed down by the target's electrons, the impacting particle undergoes nuclear elastic collisions, thus displacing atoms in the target. These first displaced atoms are called primary knock-on atoms, (PKAs).
Such PKAs act as additional self-projectiles and collide with other atoms creating a collision cascade \cite{Was2017} (in the ps regime).
Different types of defects are created as a result, such as vacancies and interstitials (Frenkel pairs), and defect clusters followed by a slow partial recombination on a longer time scale up to nanoseconds and longer.

Several physically realistic metrics for the quantification of radiation damage exist \cite{Nordlund2018,Nordlund2018B, Griffin2014}.
In the high-energy particle physics and space radiation effects communities, the Non Ionizing Energy Loss (NIEL) concept is commonly used to describe the damage due to cumulative atomic displacements.
The NIEL represents the portion of the energy of an incident particle that goes into non-ionizing phenomena in the target, being displacement damage effects. Apart from the high energy range, where non-ionizing phenomena are due to inelastic nuclear scattering, in the medium-low energy range atomic displacements nuclear stopping represents the main contribution to the NIEL, with phonons not directly leading to a displacement also eventually contributing.
On the basis of a considerable set of experimental observations, it is generally assumed that the degradation of output parameters of a semiconductor device under ion-irradiation is linearly correlated with the NIEL
\cite{8331152,srour,summ1993,mess20019,summ19942068}.
According to the commonly used 
Norgett-Robinson-Torrens model (NRT) \cite{norgett} which is based on the binary collision approximation (BCA), there is a linear relationship between the NIEL and the number of displacements being subsequently a linear function of the PKA energy.  
The NRT model is implemented in the widely used SRIM/TRIM \cite{srim} software \cite{WEB2019} and in wider-purpose Monte Carlo (MC) particle transport codes (at low energies) \cite{ALLISON2016186,fasso}.

The linearity of the NIEL means that only the number of defects should give a measure of the damage, irrespective of their nature, e.g. whether defects are clustered in small regions (as for neutron-induced damage) or homogeneously scattered over a relatively large volume (as for low-energy proton or $\gamma$-ray-induced  damage) \cite{Dawson2021}.
This independence allows to obtain the damage produced by different particles and with different energies via a NIEL scaling \cite{8331152, srour,srour2,summ1993, https://doi.org/10.1002/pip.357,summ19942068}.
Deviations from the linear dependence of the number of defects on the PKA energy were found via MD simulations on Si \cite{nord98,otto, santos1,santos2} for low energy PKAs.
These works \cite{nord98,otto, santos1,santos2} also highlighted that atomic displacements occur even for energies lower than the commonly used displacement threshold energy, implying that not all the energy transfer is to phonons below such energy but that also collective motion can generate some damage \cite{ingui2010}.
The non-linearity of the NIEL was also observed in experiments \cite{summ1993} and MC simulations \cite{etde_20318738}.
Deviations from the linearity of the degradation parameters and the NIEL were also found for different energy ranges \cite{Moll2002,SUMITA2003448}.
Efforts have been made  \cite{ingui2010,AKKERMAN2001301,akkerman2006new,Mess2003,weller,1684045}, to propose an effective or adjusted NIEL model to correct the deviations from linear dependence,
either via MD studies \cite{ingui2010, 6359810,gao2017displacement,SALS2017} or experimental works \cite{https://doi.org/10.1002/pip.357, Lu2011AdjustedNC}.

The manner in which the energy dissipation by the PKAs along the incident ion trajectory is determined and partitioned into energy loss to electrons and into atomic motion is fundamental for the counting and further evolution of defects. NIEL calculations in SRIM and wider-purposes MC codes are done without considering any coupling between ionic and electronic degrees of freedom, using amorphous targets and using a constant threshold displacement energy for each element that is independent on the material in which the element is inserted. For the nuclear stopping, the universal screening potential (Ziegler-Biersack-Littmark, ZBL) is used in SRIM, while in wider-purpose MC codes such as Geant4 \cite{ALLISON2016186} either the ZBL \cite{zieg2010} or a modified screening function given by the Moliere's approximation of the Thomas-Fermi model are used.

The electronic stopping of heavy ions is calculated via a charge scaling of the proton stopping. The latter is calculated in SRIM on the basis of fitting (and extrapolation) of experimental data \cite{WITT,WEB2019} using the (linear-response) Lindhard-Sharff model. In Geant4 this model is used for low energy protons (below $\sim$10keV/amu) only, and a parametrization based on ICRU tables and then the Bethe Bloch formula are used for increasing energies \cite{1610988}. All these approaches, however, inherit the main approximations mentioned above. 

The accuracy of such stopping power values, in particular for low energy, for self-irradiated mono- and multi-element systems, is actually unknown and still needs to be largely verified, given the paucity of available experimental data. Several recent works highlight the limitations of SRIM in the prediction of the electronic and nuclear stopping power \cite{WEB2019,crocom,WITT,Stoller2000}. 

Real Time-Time Dependent Density Functional Theory (RT-TDDFT) represents a promising parameter-free approach to obtain the electronic stopping in any target for any projectile, at energies where the assumptions of SRIM can no longer be justified. Recent studies on monoelemental systems \cite{Tamm2018,Caro2019, Tamm2019, sand2019, lee2020multiscale, jarrin2021} showed that the inclusion of electronic effects obtained via RT-TDDFT in MD cascade simulations affected both the number of defects and the cascade morphology.

In this context, the recently proposed electron-phonon (EPH) model for non-adiabatic dynamics \cite{Caro2015, CARO2017, CARO2018, Caro2019, CORREA2018, Tamm2018, Tamm2019}
improves upon the previously developed Two-Temperature model (TTM) (by Duffy \textit{et al.} \cite{Duffy2006}). 
The EPH model treats the electron-phonon coupling and electronic stopping as two manifestations of the same physical process.
\textcolor{black}{In contrast with the TTM, the friction term now depends on the local structural environment allowing for better agreement with \textit{ab-initio} electronic stopping data.}
Recent work on cascade simulations in Si showed that the EPH model correctly reproduces the density-dependent ESP and that the final radiation damage differs considerably from the TTM predictions \cite{jarrin2021}.

In this work, 
we perform an \textit{ab-initio} study of collision cascades induced by self-recoils of 100 eV, 1 keV and 10 keV, using both the EPH model and the TTM in the paradigmatic case of GaAs, whose damage is often studied in the space engineering  community being the most active layer of state-of-the-art solar cells used in space missions. We devise a more robust procedure as put forward by Caro \textit{et al.} \cite{Caro2019} to fit the EPH model to the electronic stopping determined via RT-TDDFT.
A different two-tier method was used that immediately prevents unbounded stopping and that relies on two global optimization steps to reduce the problem of local optima. 
We show for the first time that the EPH model provides a good description of the electronic stopping in the semiconductor material GaAs.
Moreover, we show that it is possible to fit the EPH model for multiple-element materials.
We additionally show that the description of the electronic stopping and electron-phonon coupling influences the number of defects, and is thus crucial to correctly model radiation damage. 
MD collision cascade simulations are performed for PKAs with different energies in GaAs and are analyzed in terms of number of defects and clustering.
How the number of defects depends on the counting criteria is also discussed.
Lastly, the number of defects is compared to the NRT model and previous works on GaAs, also discussing eventual deviations of the linearity of the number of defects with PKA energy.

\textcolor{black}{This work is part of a larger scheme as envisioned by M. Raine \textit{et al.}\cite{MRaine} spanning multiple time and length scales to model the radiation impact. Within this scheme the distribution of PKAs is initially evaluated from high energy Monte Carlo simulations, resulting in PKA energy distributions for different species. In this work the impact of the individual Ga and As PKAs is evaluated leading to damage configurations only a few picoseconds after the PKA caused the collisional cascade.
After our work, the obtained damage configurations are evolved up to a timescale of seconds using another Monte Carlo simulation (k-ART) \cite{kART}.
}

\section{Methods}

This work is divided in three sections.
Section A discusses the determination of the ESP via RT-TDDFT calculations.
Section B, the obtained ESP results are used to train the EPH model.
An iterative procedure is applied to optimize the dissipation functions that are part of the EPH model to best reproduce the \textit{ab-initio} ESP.
In the last section, the EPH model and TTM are used to run MD simulations of collision cascades.

\subsection{Calculation of the electronic stopping power}

For projectiles with high velocity, the main kinetic energy loss mechanism occurs via non-adiabatic excitations of the electronic system. This loss is usually quantified as the ESP, $S_e$ that is defined as:
\begin{equation}
    S_e(E) = -\frac{dE_e(x)}{dx}
\end{equation}
where $dE_e$ is the energy loss to the electronic system, $dx$ is the distance travelled by the projectile, i.e., the penetration depth.
We compute the ESP by simulating the passing of an ion with a constant velocity through the target material while keeping the rest of the atoms frozen, which allows to readily obtain $dE_e$ as the change of the total energy of the system (thus neglecting the nuclear stopping component). This approximation is completely valid for the projectile velocities investigated in this work \cite{Pruneda2007, Zeb2012, Ullah2018}, and it is widely employed in most of the computational works determining the ESP by RT-TDDFT methods \cite{Caro2019, lee2020multiscale, SCHL2015}. 

The RT-TDDFT calculations are performed with the plane-wave Qb@ll code \cite{Qbox, qball} using the PBE functional \cite{PBE}.
\textcolor{black}{Previous experience of calculating electronic stopping power using LDA and GGA (PBE) functionals show quite quantitative agreement with experiments \cite{lee2020multiscale, SCHL2015, CARO2017, Reeves2016}.}
Due to the high velocities of the PKAs, leading to relatively deep-level electronic excitations, pseudopotentials with 13 and 15 valence electrons are used for gallium and arsenide respectively.
The core electrons are especially relevant for highly energetic projectiles \cite{Ullah2018}.
The aim of our RT-TDDFT calculations is to parametrize the EPH model for collision cascades with PKA energies well below 100 keV.

Hamann-type norm-conserving pseudo-potentials\cite{HSC} as obtained from the quantum-simulation pseudopotential library have been used \cite{pseudos, PSGen}. 
\textcolor{black}{The pseudopotential cut-offs are 0.6 \AA\ for gallium and 0.56 \AA\ for arsenic.}
The wavefunction basis cutoff was set at 100 Rydberg and a smearing was added using 4 empty states at 1000 K.
A 2x2x3 supercell of GaAs is used for all calculations containing 96 atoms and its Brillouin zone is sampled only at the $\Gamma$-point. 

In our simulations, the wavefunctions are propagated using the enforced time reversal symmetry propagator.
In this work, the projectile, unless stated otherwise, is always neutral, i.e., it starts in an unionized form since we assume that the electronic environment of the PKA at the start of its trajectory is equivalent to that of a neutral atom in its ground state.
See Supplemental Material \textbf{S1} \cite{SM} for ESP evaluations of highly ionized As PKAs.
Below 1000 keV we used a time step of 0.4 attoseconds. At higher energies the time step is chosen such that the projectile displaces 0.005 {\AA} per simulation step.

There is an initial transient period before the projectile charge is stabilized. For this reason, only the results after this transient phase are used to obtain the average ESP. 
Every time the projectile passes close to an atom, the potential energy of the projectile shows a peak.
See Supplemental Material \textbf{S2} \cite{SM} for detailed information about how the average ESP is obtained.

\subsection{Including electronic stopping in MD}

In MD simulations, the nuclear stopping can be calculated relatively straightforwardly as it corresponds to the energy loss of the projectile due to all the interatomic collisions. However, especially at higher projectile energies 
\textcolor{black}{($E >$ 100 keV)}, the most important energy loss mechanism is the electronic stopping.
Several methods are constructed to account for the electronic stopping in MD simulations without taking the electrons into account explicitly \cite{frictionMD, Duffy2006}.
One of the earlier models applied only an extra friction force to the atoms, but the energy lost by the projectile is removed from the system and not returned as thermal energy via electron-phonon coupling.

The TTM allows both, the inclusion of the electronic stopping and the return of the energy adsorbed by the electronic system back to the nuclear system \cite{Duffy2006, Rutherford2007}.
TTM adds an effective electronic system to the classical atomic MD system with its own temperature. The normal MD equation of motion 
is exchanged by a Langevin-type equation by adding a friction term and a stochastic force term.
The friction term scales linearly with the atomic velocity (in agreement with Lindhard theory \cite{lindhard63}) and is scaled with a $\gamma_s$ parameter related to the electronic stopping.
The value of $\gamma_s$ is normally obtained from SRIM.
The energy transport in the electronic system is covered by an electronic heat equation using the electronic specific heat and electronic thermal conductivity. 

The advantage of the TTM is that it allows to include electronic stopping and electron-phonon interactions into MD simulations, however, there are a few disadvantages.
Firstly, it applies the electronic stopping such as a projectile would experience in a uniform electron gas and thus effects such as channeling are not taken into account \cite{Lohmann2020}.
Secondly, many parameters are included for which there is little experimental data, making it difficult to obtain reliable results and predictability \cite{JARRIN20201}.

Application of the TTM proved that the inclusion of the electronic temperature elongates the duration of the thermal spike observed during collision cascades, and that this enables more thermal annealing leading to decreased defect numbers. Additionally, the TTM allows to investigate radiation resistant materials that have large electron-phonon couplings \cite{Rutherford2007}.
The TTM as implemented in LAMMPS \cite{LAMMPS} only allows to give one $\gamma_s$ parameter, so for GaAs, the friction term will be the same for Ga and As projectiles.

The recently developed EPH model \cite{CARO2018, Tamm2018, Caro2019}
improves upon the TTM by modelling the electronic system by spherical electronic densities placed on each atom and thus the ESP can depend on the local electron density experienced by the projectile. 
The electron-phonon coupling is treated as an electronic stopping process at low energies (meV scale).
Thirdly, it was proven that TDDFT can give accurate results for the ESP and the electron-phonon coupling \cite{Ullah2018, Caro2015} making it possible to parametrize the EPH model using TDDFT results.
In \cite{Caro2019}, it is shown that ESP profiles from a well fitted EPH model closely match the RT-TDDFT results.

In the EPH model, the friction and random force are described by many-body tensors, contrary to the TTM in which they are scalar \cite{Tamm2018}:
\begin{equation}\label{eq:eph1}
    m_I\frac{\partial \mathbf{v}_I}{\partial t} = \mathbf{F}_I 
- \sum_J B_{IJ} \mathbf{v}_J
+ \sum_J W_{IJ} \boldsymbol\xi_J
\end{equation}
The second and third terms on the right-hand side correspond to the friction force and to the random force acting on the $I^{th}$ atom.
The forces acting on the $I^{th}$ atom now depend on the velocity of all other atoms $J$, and thus all forces are correlated.
The set of $\boldsymbol\xi_J$ are white-noise mutually-uncorrelated Gaussian random variables normalized
to $2k_B T_e$ \cite{Caro2019}.
$B_{IJ}$ and $W_{IJ}$ are two tensors that describe the spatial correlations and they are related via the fluctuation-dissipation theorem such that:
\begin{equation}\label{eq:eph2}
    B_{IJ} = \sum_K W_{IK} W_{JK}^{\text{T}}
\end{equation}
Both the ESP and the electron-phonon coupling are determined by the terms in the $W$ matrix.

The terms in $W$ are constructed such that the random forces do not produce any local net translational nor rotational momentum. This condition is motivated by the preservation of the correct phonon lifetimes and polarization. $W$ is then defined as:
\begin{equation}\label{eq:eph3}
    W_{IJ} =
    \left\{
        \begin{aligned}
            &-\alpha_J(\bar{\rho_J})\frac{\rho_I(r_{IJ})}{\bar{\rho_J}}\vec{e}_{IJ} \times \vec{e}_{IJ} &\;\; (I \ne J)\\
            &\alpha_I(\bar{\rho_I})\sum\limits_{K \ne I}\frac{\rho_K(r_{IK})}{\bar{\rho_I}}\vec{e}_{IK} \times \vec{e}_{IK} &\;\; (I = J) \\
        \end{aligned}
    \right.
\end{equation}
where $\rho_I(r)$ is the value of the spherical density $\rho$ (centered at atom $I$) at a distance $r$.
$\bar{\rho_I} = \sum_{J\neq I} \rho_J(r_{IJ})$ is the electron density experienced by atom $I$ and is the sum of contributions from all its neighbors.
$\alpha_I(\bar{\rho_I})$ is a function defining the coupling between the electrons and ions \cite{Tamm2019}.

The shape of the $\alpha$ function is free and depends on the atomic species. Thus for every element a different $\alpha$ function is used. 
The $\alpha$ function has to be defined over the whole relevant range, from low densities occurring at crystal vacancies and in the center of wide channel directions, to high densities encountered at close collisions.
This function is optimized using the RT-TDDFT data. The ESP has to be defined in a wide range of electronic densities in order to obtain an accurate $\alpha$ function along the whole density range.

The electronic heat equation is very similar to the TTM and the same values for $C_e$ and $\kappa_e$ are used \cite{Tamm2019}.
The EPH model is available as an external plug-in for LAMMPS.

\subsection{Collision Cascade MD simulations}

Following the approach of Gao \textit{et al.} \cite{gao2017displacement}, we use the bond order potential as developed for GaAs by Albe \textit{et al}. \cite{Albe2002} combined with the ZBL functional \cite{ziegler1985stopping} at small distances to describe the short range repulsive interactions that occur during strong collision impacts. (See Supplemental Material \textbf{S3} \cite{SM} for the potential visualization.)

\textcolor{black}{For all MD simulations, periodic boundary conditions are applied. The total simulation box is a large cubic supercell. All the outer cells of the simulation supercell form a thermostat keeping the atomic temperature at 300 K via a Nose-Hoover thermostat \cite{Shinoda}, thus absorbing the thermal wave. The EPH or TTM equations only apply within the interior region.}

For each simulation type, the system is first initiated at 300 K and equilibrated for a sufficient time of 100 ps. Afterwards, the equilibrated system is used as a starting point for a set of collision cascades.
At the beginning of the cascade, the time step is set to 0.001 fs. Subsequently, the time step is updated such that no atom moves more than 0.05 \AA\ per time step.
All simulations are run for several picoseconds until the temperature returns back to 301 K, so the thermal spike is sufficiently dissipated. 
\textcolor{black}{This allows us to compare the direct effect of using different simulation types but for experimental comparison longer time scale methods would be required to include longer timescale effects.}

To get a good statistical average, for each MD setting, 78 different cascades are run with different initial PKA directions chosen according to the procedure outlined in reference \cite{JARRIN20201}.
The PKA is chosen such that its initial velocity vector points as much as possible towards the center of the interior region to keep the collision damage into the box and preventing atoms from leaving the interior during the cascade.
The clustering analysis is done by using the Louvain algorithm for community detection \cite{Blondel_2008}.

\section{Results and discussion}

\subsection{Calculation of the electronic stopping power}

\subsubsection{Choosing the PKA directions}

Different crystal directions are chosen for the PKA trajectories such that a variety of crystal environments and electron densities are probed to enable accurate subsequent fitting of the EPH model.
We included three full channeling directions where the projectile travels in the middle of the channel, thus traversing a lowest density path through the crystal: $\langle001\rangle$, $\langle011\rangle$ and $\langle111\rangle$.

\begin{figure}[htbp]
\centering
\noindent\includegraphics[width=.15\textwidth]{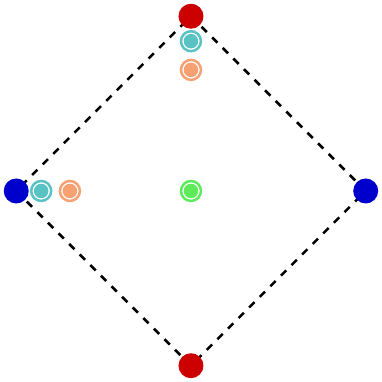}
\caption{Schematic representation of the $\langle001\rangle$ channel. Blue dots signify arsenic and red dots signify gallium atoms. 
The green dot indicates the center channeling direction $\langle001\rangle$.
The turquoise dots indicate the position of the $\langle001\rangle_{Ga}^{7}$ and $\langle001\rangle_{As}^{7}$ off-center channeling directions, located at one seventh of the atom-center distance.
The orange dots indicate the position of the $\langle001\rangle_{Ga}^{4}$ and $\langle001\rangle_{As}^{4}$ off-center channeling directions, located at a quarter of the atom-center distance. 
}
\label{fchannels}
\end{figure}

Off-center channeling directions, i.e., directions parallel to the channel but closer to the atoms, are added in the $\langle001\rangle$ channel.
We adopt the following naming scheme: $\langle 001\rangle_{X}^{N}$ indicates a trajectory moved away from the $\langle001\rangle$ center towards the atoms of type $X$, located at $\frac{1}{N}$ of the atom-center distance. For the included off-center directions see Figure \ref{fchannels}.

Random trajectories that are incommensurate with any crystallographic symmetry vector are also included. 
Since these directions are not periodic, they are selected to probe each possible density environment of the crystal.
These so-called off-channel directions are chosen using several conditions:
\begin{itemize}
    \setlength{\itemsep}{1pt}
    \setlength{\parskip}{1pt}
    \setlength{\parsep}{1pt}
    \item The starting point of the PKA has to be not too close to another atom to get a good starting wavefunction.
    \item Due to the choice of a $2\times2\times3$ supercell, the PKA direction has to have the largest velocity component in the $z$ direction.
    \item Since the off-channel directions are sampled for a longer trajectory, the projectile has to re-enter the unit-cell as far as possible from where it has already been before.
    \item The trajectory should not contain a too close encounter (frontal collision) with another atom.
    \item The trajectory should be as incommensurate with a crystallographic direction as possible. To test this, we use the auto-correlation function, which tests if there is any periodicity in a signal. We record the closest distance to another atom as a function of the trajectory along the first 60 \AA. If there are no large peaks (except the self correlation) in the auto-correlation function, the chosen direction is sufficiently incommensurate with any low-index crystallographic direction.
\end{itemize}

\textcolor{black}{
In the EPH model, the electron-phonon interactions are treated as an electronic stopping process that happens around the atomic reference positions. The electronic stopping at these (low atomic density) reference positions is sampled by including an incommensurate PKA trajectory moving through a vacancy.}

\subsubsection{Electronic stopping power calculations}

\begin{figure}[htbp]
\includegraphics[width=\columnwidth,
]{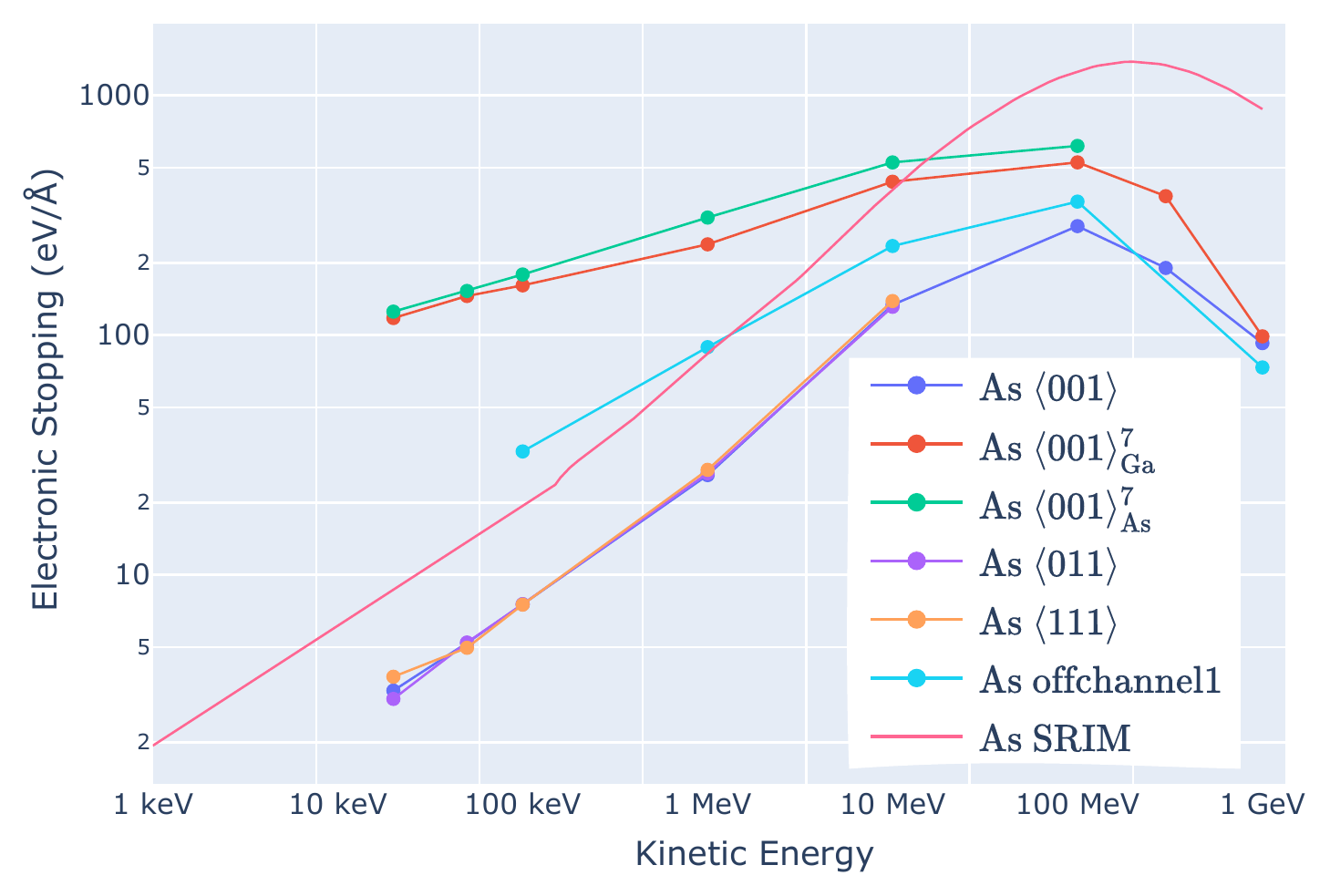}
\caption{Electronic stopping of Arsenic PKAs in Gallium Arsenide. (Some points are omitted to cut the computational cost.)}\label{fAsSe}%
\end{figure}

The ESP for several trajectories was evaluated for a range of different energies. Since the results for gallium and arsenic PKAs are qualitatively similar, in Figure \ref{fAsSe} we show only the ESP for arsenic PKAs for simplicity. Also the stopping power as obtained by SRIM is included.

The ESP values for all channeling directions are smaller than the SRIM results, as is also observed for many other materials \cite{SCHL2015, natalia, PhysRevB.91.125203, Ullah2018}.
We observe that the ESPs along the centers of the three channels are very similar.
For example, for 1 MeV As PKAs, the ESP values are 26.0, 26.5 and 27.4 eV/\AA\ for the $\langle001\rangle$, $\langle011\rangle$ and $\langle111\rangle$ channels respectively.
This deviates from other works on materials with a similar diamond-like structure \cite{natalia, PhysRevB.91.125203} where the wider $\langle011\rangle$ channel resulted in slightly lower ESP values than the $\langle001\rangle$ and $\langle111\rangle$ results. 

The ESP results for the off-center channeling close to the atoms, have a very high ESP due to the high electron densities encountered during the projectile trajectory. 
The ESP for $\langle 001 \rangle_{\text{As}}^7$ is slightly higher than for the $\langle 001 \rangle_{\text{Ga}}^7$,
so the stopping on the projectile passing an As atom is higher, in agreement with its higher electron density.

The off-channeling results are in between the channeling and off-center channeling results. This is expected as the off-channeling trajectory probes the electron densities most evenly (time averaged) while channeling and off-center channeling mainly probe either only low or mainly high electron densities.

Since SRIM does not take into account channeling, i.e., it considers only a series of binary collisions, the off-channeling result should match best with the SRIM result. 
We confirm this assumption at lower energies ($\leq$ 10 MeV).
At higher projectile energies, there is a larger deviation from SRIM. 
This is also observed in several other works \cite{Ullah2018, lee2020multiscale} and can be attributed to the increasing importance of core electrons at higher projectile velocities.
Our calculations only explicitly treat 13/15 valence electrons of the 31/33 electrons of Ga and As respectively so the effect of core electrons is not included.

To fit the EPH model, we focus on the results at the PKA energy of 100 keV, since it offers the best balance between accuracy and computational time, as going to lower energies (such as 20 keV and lower) drastically increases the computational costs while at much higher energies, the linear relationship between the ESP and the PKA velocity breaks down \cite{Duffy2006}.

\begin{figure}[htbp]
\centering
\noindent\includegraphics[width=.5\textwidth]{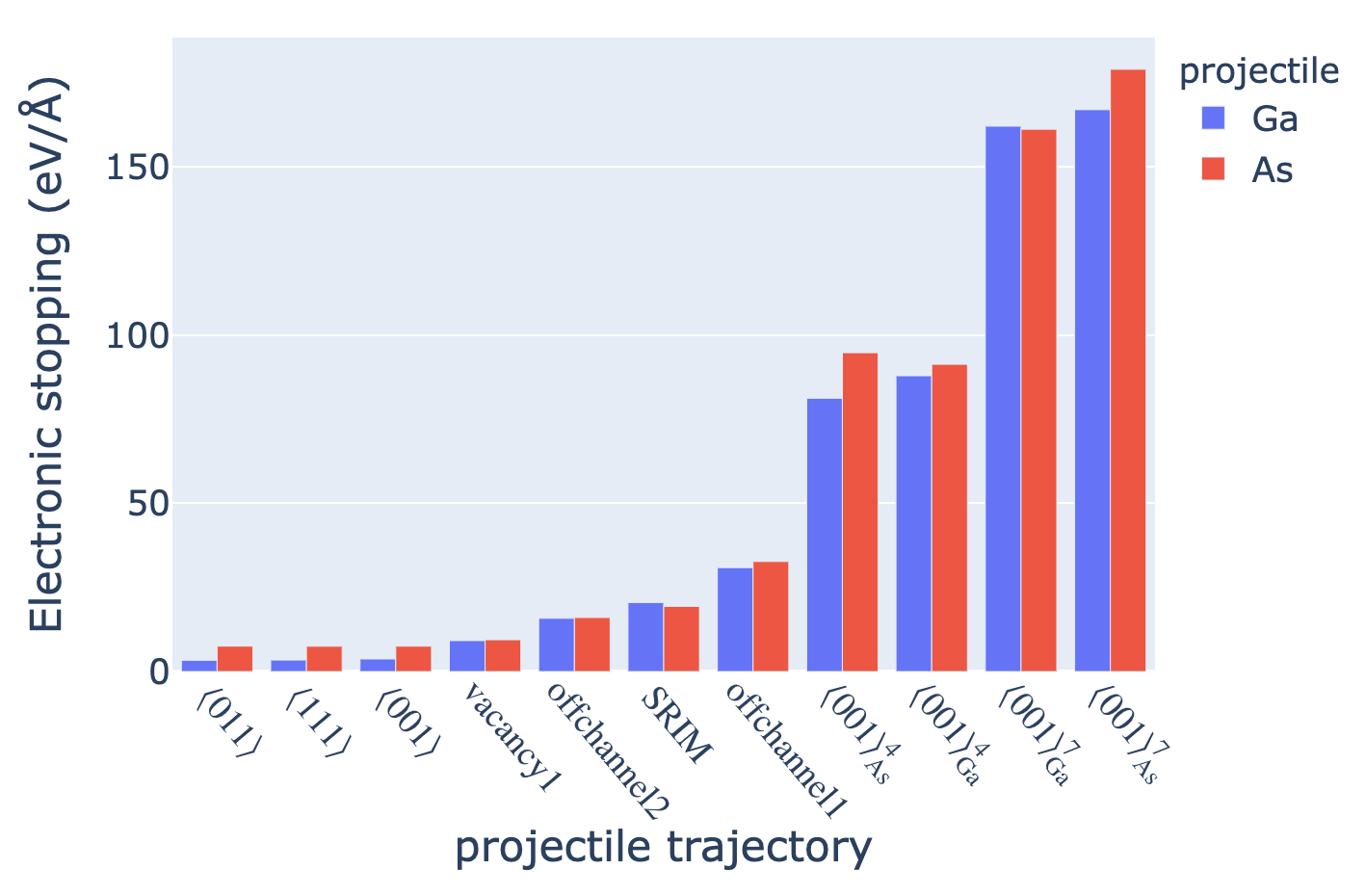}
\caption{Electronic stopping values of Ga and As PKAs in GaAs along different trajectories with PKA energy of 100 keV. 
}
\label{ftraj}
\end{figure}

Figure \ref{ftraj} shows the ESP values at multiple trajectories both for Ga and As PKAs, all with the kinetic energy of 100 keV. Some additional trajectories are considered such as $\langle001\rangle_{Ga}^{4}$ and $\langle001\rangle_{As}^{4}$.
Another off-channeling trajectory is included that does not exhibit a close collision as does the \textit{offchannel1} direction around 38 \AA\ (See Supplemental Material \textbf{S2} \cite{SM}).
Last, an off-channeling trajectory moving through a vacancy is included (\textit{vacancy1}). 

The ESPs along the off-center channeling directions $\langle001\rangle_{X}^{4}$ are lower than along the $\langle001\rangle_{X}^{7}$ trajectories (with $X$ being Ga or As).
The ESP along the second off-channel direction without a close collision is clearly lower than the on on the first off-channeling direction and is very close to the SRIM result.
The trajectory moving through a vacancy yields a lower ESP than along the other off-channeling directions.

In general, the differences in ESP between Ga and As are small and most pronounced for the channeling directions.
The ESP for As is typically larger than for Ga.
Off-center channeling trajectories close to the As atoms have also higher ESPs than those close to the Ga atoms, in accordance with the higher electron density around As.

\subsection{Fit of the EPH model to the ESP results}

Using the set of ESP results along the different trajectories, we fit the EPH model to include the ESP into the MD simulations. 
Since it would be too costly to run a self-consistent quantum-mechanical calculation within each MD step, in the EPH model the electron density is approximated as the sum of the spherical densities centered on each atom. Thus, one needs to provide the spherical densities of Ga and As. These atom-like densities are obtained using the OPIUM program on the isolated atoms \cite{opium1} (See Supplemental Material \textbf{S4} \cite{SM} for the atomic density visualisations).

\subsubsection{Optimizing the $\beta$ function}

The $B$ matrix in Equation \ref{eq:eph1} represents the friction coefficient and is the Hermitian square of $W$ (Equation \ref{eq:eph2}). For this reason, the friction scales with $\alpha^2$. 
In the following section, we optimize this $\alpha^2$ function, which we call $\beta$, following the same notation as in reference \cite{Caro2019}.

\begin{figure*}[tb]
\centering
\noindent\includegraphics[width=\textwidth]{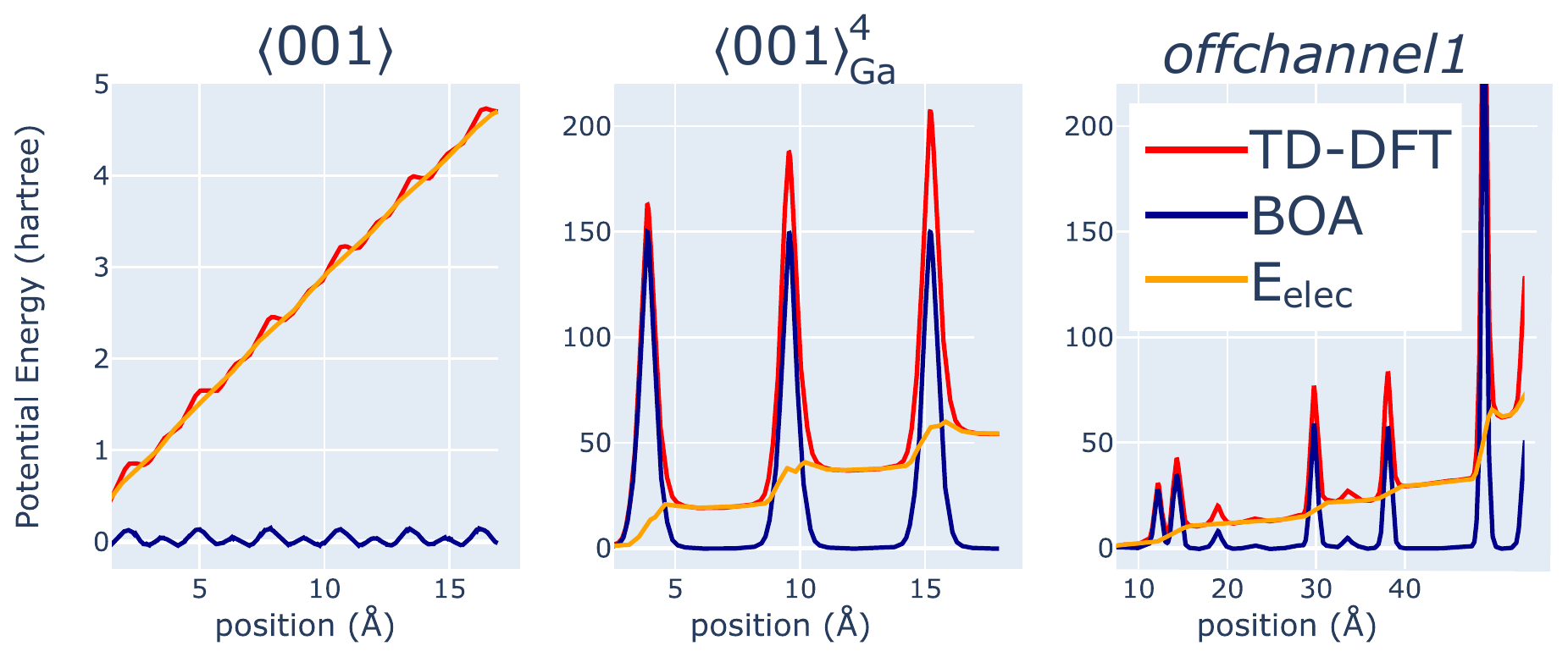}
\caption{Three examples of how the instantaneous dissipation energy (orange) due to electronic stopping is obtained by subtracting the BOA energy (black) from the RT-TDDFT (red). Here, they are shown for Arsenic PKAs of 100 keV, 
along the $\langle001\rangle$, $\langle001\rangle_{Ga}^{4}$ and \textit{offchannel1} trajectories.
}
\label{fTDBOA}
\end{figure*}

In principle, $\beta$ depends not only on the local density, but also on the directions of the moving particles.
Especially at higher densities, it can be observed 
(see especially at the peaks of the middle plot of Figure \ref{fTDBOA})
that at the beginning of a close collision, when two atoms get closer to each other, the ESP is high, while further on, when the particles move away from each other, the ESP is lower. Nevertheless, the time-independent density profile is symmetric. This means that the ESP cannot be expressed exactly as a function that only depends on the time-independent density \cite{Caro2019}.
It is thus an approximation of the EPH model to let $\beta$ be only a function of the time-independent density, approximated by the sum of the spherical densities of each atom called $\rho_0$. 

In the following section, we try to find a $\beta(\rho_0)$ function. 
We define that the best $\beta$ function is the one that, when the same trajectories are run with EPH-MD and RT-TDDFT, gives the smallest difference in energy dissipation due to the ESP.

From the EPH-MD simulations, we directly obtain the energy dissipated by the system due to the ESP at each point along the trajectory. 
To obtain the instantaneous dissipation, or instantaneous ESP, we have to subtract the effect of the crystal structure from the RT-TDDFT energy curves. 
The structure effect can be removed by computing the energy corresponding to the ground state (BOA) of each atomic configuration, and then subtracting it from the energy obtained during the RT-TDDFT run at the same configuration:
Now, the energy increase caused by the non-adiabatic effects only, $E_{elec}(r)$, which equals the dissipated energy to electronic stopping, is obtained as:
\begin{equation}
    E_{elec}(r) = E_{RT-TDDFT}(r) - E_{BOA}(r)
\end{equation}

In Figure \ref{fTDBOA}, we show for three trajectories how the effect of the structure is removed to get $E_{elec}$. 
For the $\langle001\rangle$, the $E_{elec}$ curve is very smooth, so the ESP is very stable along the channel. From the other two trajectories, it can be observed that most energy is dissipated during close collisions.

We choose to define the error for a given $\beta$ function as the difference between the EPH-MD and RT-TDDFT dissipation. For a single trajectory the mean absolute error is defined as:
\begin{equation}
    MAE = \frac{1}{r_{12}}\int_{r_1}^{r_2} \left| E_{MD}(r) - E_{elec}(r) \right| dr
\end{equation}
And the total error to be minimized is the sum of all $MAE$s from all the trajectories taken into account during the minimization.

There are different strategies to solve the challenge of finding a $\beta$ function that minimizes the total $MAE$.
Generally, we assume that when there are no electrons, there is no electronic stopping so the function has to pass through the origin. 
We also expect the ESP to increase when the density increases, but only until a certain density is reached after which the ESP either saturates, or decreases.
Different strategies can be applied. Caro \textit{et al}. \cite{Caro2019} constructed a $\beta$ function by using a cubic spline with 6 knots and afterwards added an exponential decay to prevent unbounded stopping at high electron densities.
Jarrin \textit{et al}. \cite{jarrin2021} used a function of the form: $\beta(\rho)= c_1\rho\cdot e^{c_2(\rho-c_3)}$ where the $c$'s are optimization parameters.

Both approaches have their disadvantages. The optimization space using the splines is non-linear so the optimization problem is prone to find a local optimum while the functional form is too restricted.
We decided to combine both approaches. 
First, a simple functional form is optimized to get a general shape of the $\beta$ function.
Subsequently, a second optimization is run that adds a cubic spline function onto the earlier function. This second optimization essentially fine-tunes the first function by making small adjustments to the function but keeps the general function shape.
The advantage of the two-tier optimization procedure is the reduced dimensionality of the problem.
In both optimization steps, the DIRECT global derivative-free optimization algorithm \cite{GN_DIRECT} is used as implemented in the non-linear optimization software NLOPT \cite{NLOPT}.
During the optimization procedures in each iteration, NLOPT proposes a trial $\beta$ function. Next, short LAMMPS calculations are run for all the relevant trajectories using the trial $\beta$ function. Subsequently, the total error is evaluated.

\begin{figure}[htbp]
\centering
\noindent\includegraphics[width=0.5\textwidth]{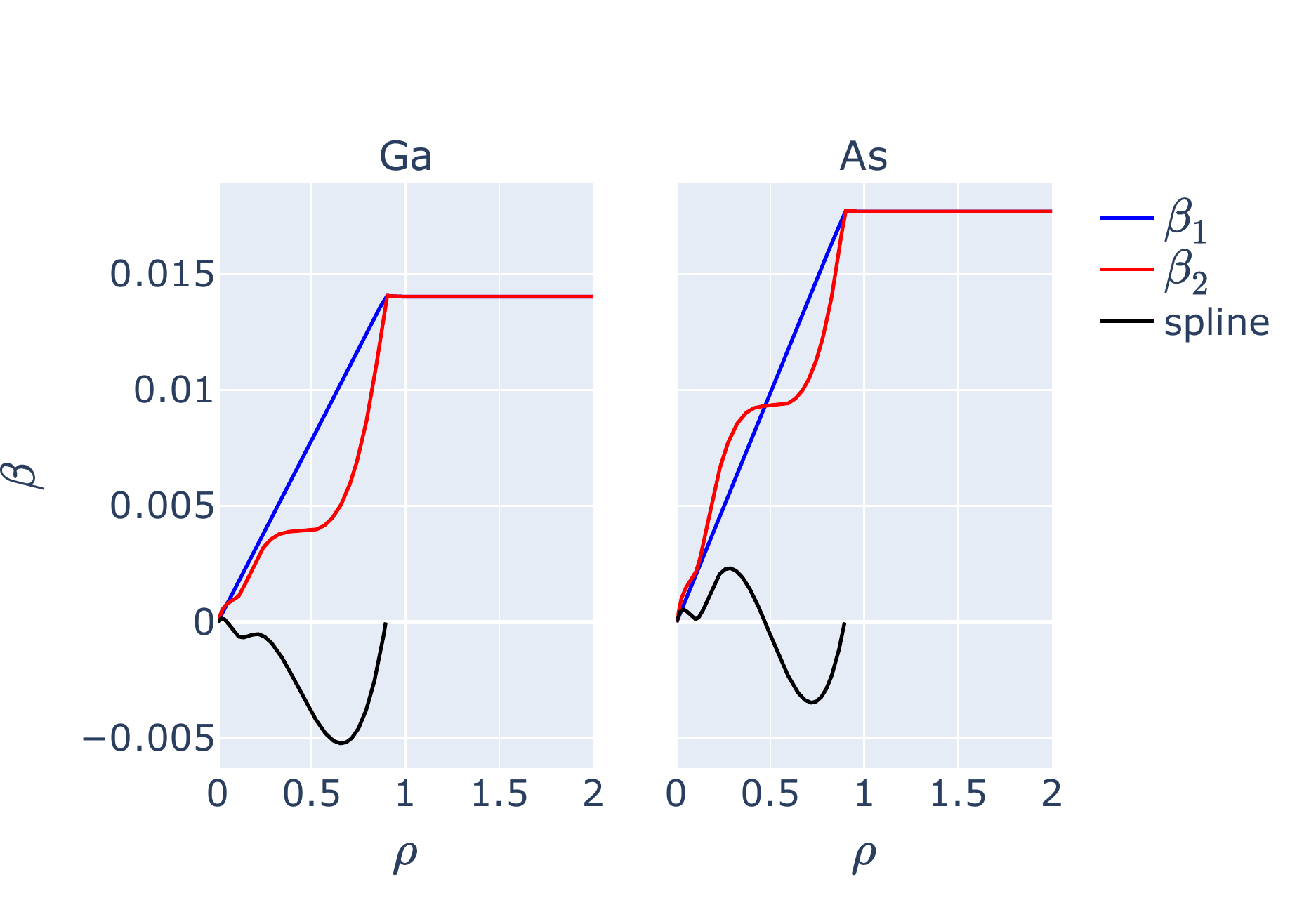}
\caption{The optimized $\beta (\rho)$ functions for gallium and arsenic with the density in $e^-/$\AA$^3$.
The $\beta_1$ curves are the result of the first optimization step, following Equation \ref{eq:beta1}. The $\beta_2$ curves are the result of the sum of $\beta_1$ and the splines correction curves. $\beta$ values are in eV$\cdot$ps/\AA$^2$.
}
\label{fbetas}
\end{figure}

Note that in our case, for GaAs, we have two elements and so two $\beta$ functions are optimized simultaneously.
As we noted that at high densities the ESP is smaller but does not drop to zero, instead of the exponential decay we decided to saturate to a constant $\beta$ function at higher densities. The use of a constant $\beta$ function is also used in \cite{jarrin2021} and \cite{Tamm2018}.
\textcolor{black}{(The trajectories 
$\langle 001 \rangle_{\text{Ga}}^7$ and $\langle 001 \rangle_{\text{As}}^7$
were not used during the EPH model training as the high density regions sampled along these trajectories are rarely encountered during the collisional cascades and the dissipation functions will be constant at these densities.}
In the first optimization step, we use the simple functional form:
\begin{equation}
    \beta(\rho) =
    \left\{
        \begin{aligned}
            & c_1\cdot \rho & \;\; (\rho < c_2) \\
            & c_1\cdot c_2 & \;\; (\rho > c_2) \label{eq:beta1}\\
        \end{aligned}
    \right.
\end{equation}
In the second optimization step, a 4-point cubic spline is added in the range [0, $c_2$] that has to be zero at both ends of the range.
In Figure \ref{fbetas}, the optimized $\beta$ functions are displayed. 

\begin{figure*}[htbp]
\centering
\noindent\includegraphics[width=\textwidth,height=7.6cm]{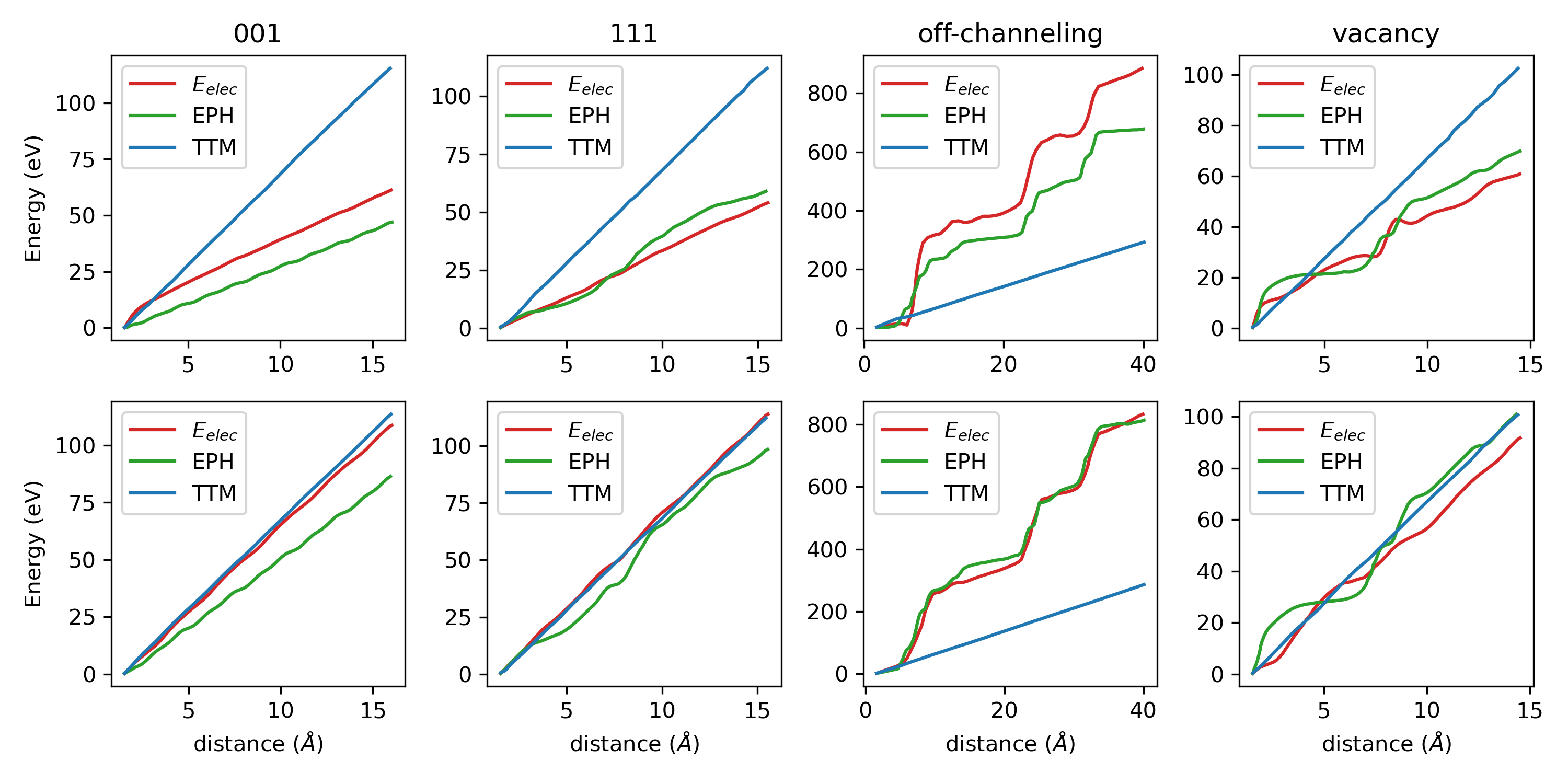}
\caption{
Comparison of the dissipation energy due to electronic stopping evaluated via RT-TDDFT ($E_{elec}$), EPH-MD and TTM-MD \textit{vs}.\ distance along the trajectory. Upper row for gallium PKAs and lower row for arsenic PKAs along the indicated trajectories. 
}
\label{fEPHcurves}
\end{figure*}

In Figure \ref{fEPHcurves}, the comparison of the EPH-MD and TTM-MD runs with the \textit{ab-initio} results are shown for 4 trajectories for Ga (top) and As (bottom) PKAs.
The TTM result has significant errors along most trajectories, as it does not take into account any local crystal information.
The EPH model is clearly more in sync with the RT-TDDFT data, although the agreement could certainly be improved by using more advanced fitting methods.

\textcolor{black}{
We decided to optimise (using NLOPT) the $\gamma_s$ parameters of the TTM using the same error measure as for the EPH model to get a fair comparison with the EPH model.
The optimized $\gamma_s$ parameters are 45.82 and 43.34 g/mol/ps for Ga and As respectively.
}
The SRIM obtained $\gamma_s$ values are 54.54 and 51.51 g/mol/ps for Ga and As, respectively. These values are clearly higher than our RT-TDDFT based values and thus would result in a too high ESP.

The total MAE for EPH is 1421 eV while for TTM it is 4680 eV. Thus with the EPH as fitted here, a 70\% reduction in the dissipation error with respect to the TTM is obtained.

\subsection{Collision Cascade MD simulations}

In this section, we present the collision cascades obtained using the TTM and EPH model
\textcolor{black}{as well as cascades without electronic stopping.}
In both models we used specific heat, $C_e$ and thermal conductivity, $\kappa_e$ values of $6.45 \cdot 10^{-6}$ eV/K and $1.0 \cdot 10^{-9}$ eV/K/\AA/ps respectively.
As shown by Jarrin \textit{et al} \cite{jarrin2021}, the EPH-MD results do not strongly depend on these electronic parameters, while for the TTM-MD results, only changing $C_e$ had a large impact.
\textcolor{black}{In the TTM model, the critical velocity was set to 96.28 \AA/ps}

We ran MD simulations of collision cascades with PKA energies of 100 eV, 1 keV and 10 keV. The MD systems were $N*N*N$ supercells with $N$ being 20, 30 or 50 conventional GaAs unit cells respectively.
Above 10 keV, the cascade structures become self-similar due to subcascade formation \cite{Nordlund2001GaAs}.

\begin{figure}[htbp]
\centering
\noindent\includegraphics[width=\columnwidth]{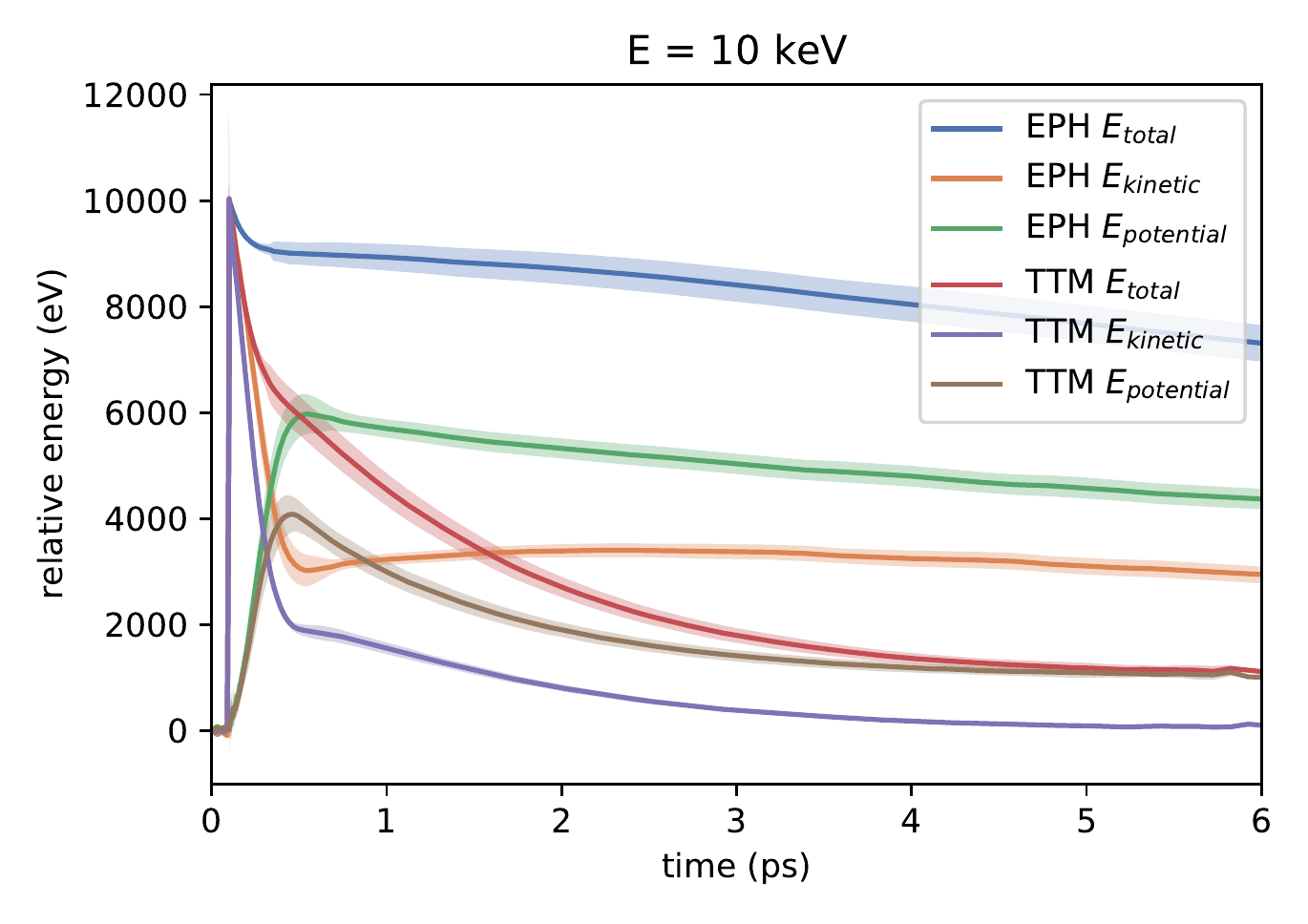}
\caption{Average evolution of the energies indicated in the figure during the MD simulations. Each curve is the average of 78 MD runs and the shaded area the standard deviation.}
\label{fMDcurves}
\end{figure}

In Figure \ref{fMDcurves}, we show how the energies evolve over time during the collision cascade.
All curves start with a peak of kinetic energy due to the projectile's direct transfer of kinetic energy followed by a peak of the potential energy, the so-called heat spike. 
The dissipation of the heat is much faster with TTM-MD than with EPH-MD.
In the evolution of the kinetic energy in the EPH-MD simulations, we can observe a second slight increase around 2-3 ps, caused by the the electron-phonon coupling, where the hot electrons dissipate heat to the ions, thus increasing again the kinetic energy of the system.
As we ran the cascade until the temperature was back at 301 Kelvin after the heat spike, running times for EPH-MD were significantly longer than for TTM-MD. 
The average run time for the MD calculations with PKA energies of 10 keV were 4 picoseconds for TTM-MD \textit{vs}.\ 51 picoseconds for EPH-MD. 

\begin{figure*}%
    \includegraphics[width=\textwidth]{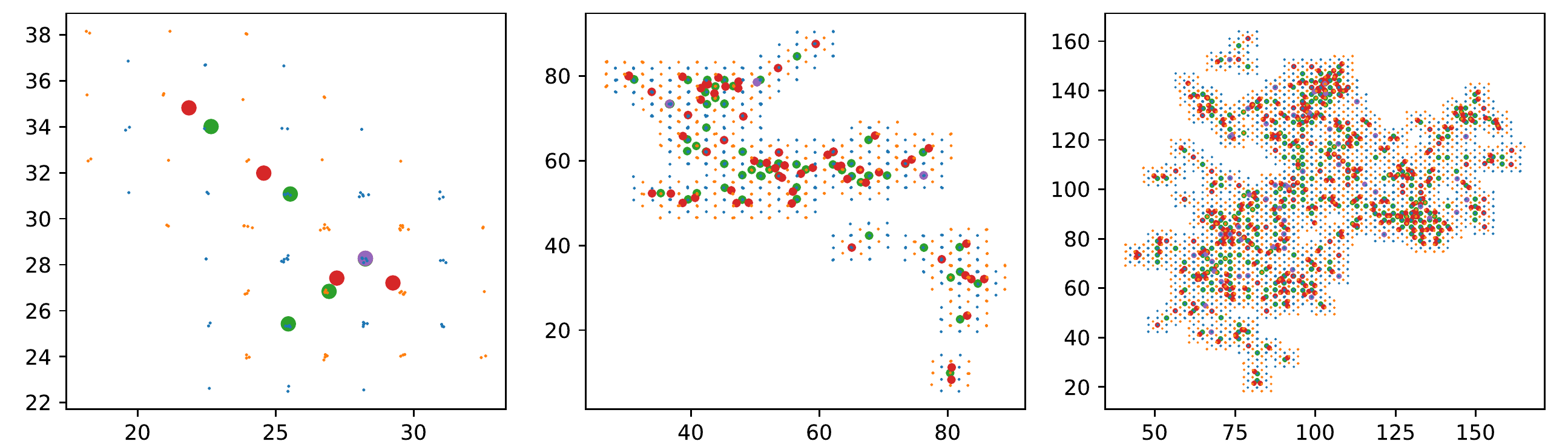}
    \caption{Three examples of final damage configurations shown in the $xy$-plane, ran with EPH-MD with different Ga PKA energies of 100 eV, 1 keV and 10 keV respectively. Insterstitials are red, vacancies green and antisites are purple. The axes dimensions are in \AA. The original, non-defect crystal atoms are only shown in proximity of the defects and depicted as small dots: gallium (blue) and arsenic (orange).}%
    \label{fig:examples}%
\end{figure*}

There are two main methods to evaluate the number of defects.
\begin{itemize}
    \item The Wigner-Seitz method \cite{WS}: The cascade is initiated on a perfect lattice. After the cascade, every atom is assigned to its closest reference crystal position. Each reference position that has two or more atoms for which it is the closest position is counted as an interstitial. Every reference position that has no atom for which it is the closest lattice position is a vacancy. 
    Antisites are defined as reference positions with only one affiliated atom but of another type as the original type at that reference position.
    \item Cut-off methods \cite{Hensel1998}: A sphere is centered on each original crystal position. After the cascade, each sphere that is empty is labelled a vacancy while each atom that is located outside any sphere is an interstitial.
    Antisites are defined as spheres with the cut-off radius centered on the original crystal atoms that have one atom within the sphere that is of a different element than the original element.
\end{itemize}
Here, we choose to use the second method with a cutoff of 1.0 \AA, which is within the stable regime where the number of defects w.r.t. the cutoff radius is most stable.
See Supplemental Material \textbf{S5} \cite{SM} for how the number of defects changes w.r.t. the defect definition and cut-off value.
The number of defects and antisites determined via the cut-off method as well as the cluster sizes and the PKA penetration depths are shown in Table \ref{tabMD}.
Three typical final damage configurations obtained with EPH-MD for the Ga PKAs are presented in Figure \ref{fig:examples}.
As expected, higher PKA energies lead to higher numbers of defects, antisites clusters and PKA depths.

The number of defects and antisites determined via the cut-off method as well as the cluster sizes and the PKA penetration depths are shown in Table \ref{tabMD}.
The results of the EPH-MD simulations for As and Ga PKAs are similar.
The average PKA penetration depths and number of clusters formed is only slightly higher for Ga than As, being always within the respective error bars of each other.
In the EPH-MD runs without electronic stopping, the number of defects is around 10\% higher 
(see Table \ref{tabMD}), so the electronic stopping effectively removes energy from the nuclei, resulting in lower final radiation damage.

A higher number of defects, antisites, and clusters is obtained for MD with the EPH model as compared to TTM.
This indicates that with the TTM, more energy is dissipated thanks to the ESP and thus as heat, while with EPH model, since there are regions with lower ESP (channeling-like conditions), more energy goes into nuclear stopping, leading to more defects. 
The higher dissipation to electronic stopping in the TTM-MD case goes in the same direction as in SRIM, which also overestimates the ESP \cite{mess2001, crocom}.
\textcolor{black}{
The SRIM obtained values for $\gamma_s$ are higher.
They would also be higher when we would have optimized $\gamma_s$ using only a random off-channeling trajectory $\gamma_s$.
Running TTM-MD with higher $\gamma_s$ values would result in a stronger ESP and the number of defects would be even lower.
}
Also, our $C_e$ and $\kappa_e$ values are relatively low, but using higher values would presumably lead to even lower number of defects, in analogy with reference \cite{jarrin2021}.
The number of clusters from the TTM-MD is clearly lower, even when taking into account the lower amount of defects; an indication that TTM-MD leads to larger defect clusters. This behaviour is also observed for Si \cite{jarrin2021}.

\begin{table*}[htb]
\caption{Number of defects and antisites as determined via the cut-off method ($1$ \AA) as well as cluster sizes} and PKA depths for the MD calculations for specified PKA and initial PKA energies. 
All values are mean values of 78 MD simulations. Values in between brackets are standard error of the mean (SEM) values.
\label{tabMD}
\begin{tabularx}{\textwidth}{@{\extracolsep{4pt}} llXXXX @{}}
\hline
                                   &             & \multicolumn{2}{c}{EPH-MD}    
                                   & NVE   
                                   & TTM-MD        \\ 
                                   \cline{3-4} \cline{5-5} \cline{6-6}
                                   & $E_0^{PKA}$ & As          & Ga           & Ga           & Ga         \\ \hline
\multirow{3}{*}{N defects}         & 100 eV      & 6 (2)       & 7 (2)        & -            & 6 (2)      \\
                                   & 1 keV       & 59 (11)     & 54 (8)       & 61 (8)       & 52 (8)     \\
                                   & 10 keV      & 542 (86)    & 542 (93)     & 592 (111)    & 464 (52)   \\ \hline
\multirow{3}{*}{N antisites}       & 100 eV      & 0 (0.5)     & 0 (0.5)      & -            & 0 (0.4)    \\
                                   & 1 keV       & 5 (2)       & 5 (2)        & 5 (2)        & 2 (2)      \\
                                   & 10 keV      & 64 (15)     & 64 (13)      & -            & 26 (6)     \\ \hline
\multirow{3}{*}{N clusters}        & 100 eV      & 2.9 (0.7)   & 3.6 (1.0)    & -            & 3.4 (0.8)  \\
                                   & 1 keV       & 14.8 (3.0)  & 16.3 (2.8)   & 16.7 (3.1)   & 13.0 (2.6) \\
                                   & 10 keV      & 96.3 (13.0) & 100.1 (13.3) & 106.3 (14.3) & 67.3 (9.2) \\ \hline
\multirow{3}{*}{PKA depth (\AA)}   & 100 eV      & 4 (3)       & 8 (4)        & -            & 7 (3)      \\
                                   & 1 keV       & 21 (12)     & 29 (16)      & 31 (17)      & 24 (12)    \\
                                   & 10 keV      & 81 (59)     & 89 (55)      & 112 (69)     & 98 (64)    \\ \hline
\end{tabularx}
\end{table*}

\begin{figure}[htbp]
\centering
\noindent\includegraphics[width=\columnwidth]{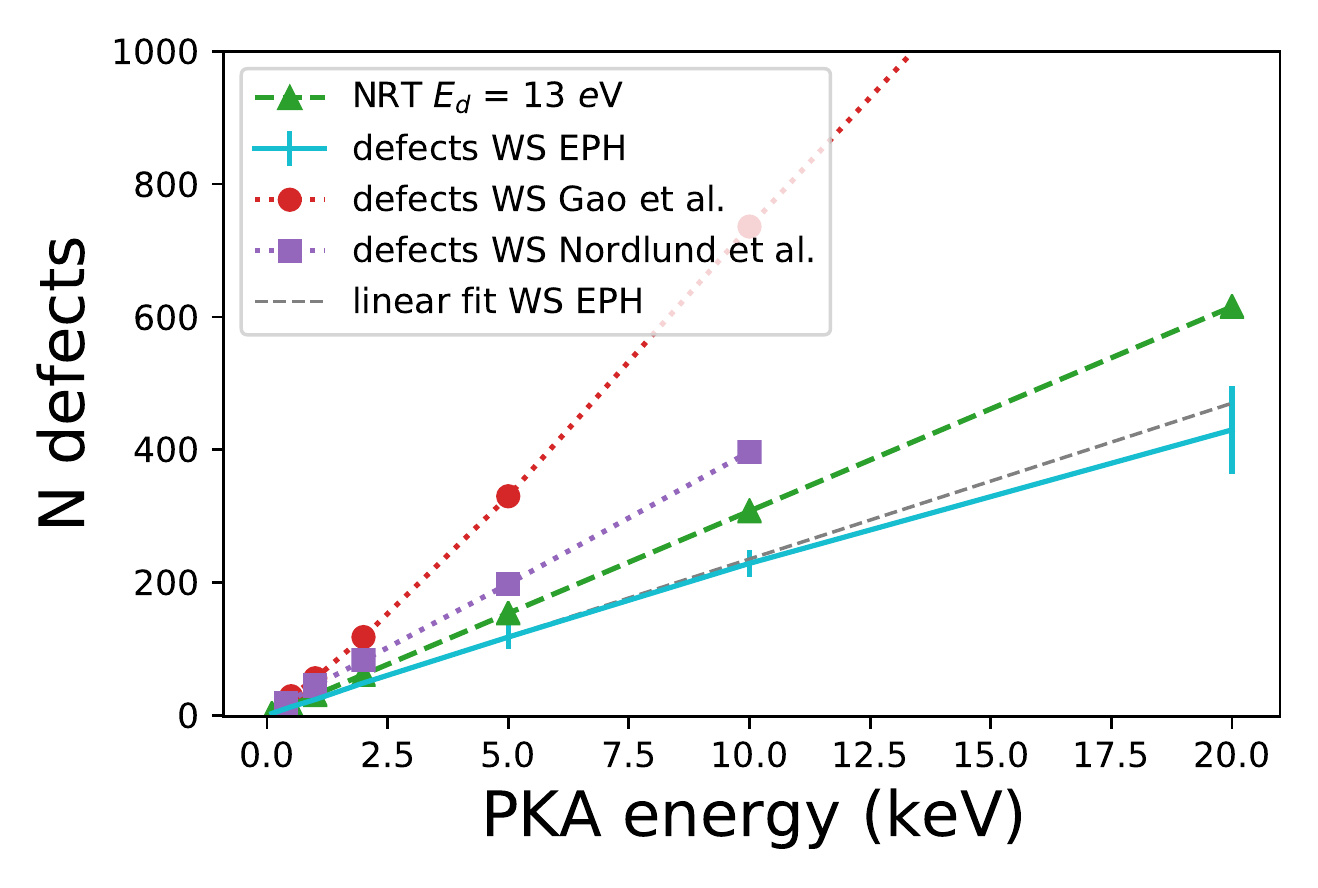}
\caption{
Number of defects for multiple PKA energies. 
Number of frenkel pair defects as determined via the Wigner-Seitz (WS) method are compared with the numbers reported in Gao \textit{et al.} \cite{gao2017displacement} and Nordlund \textit{et al.} \cite{Nordlund2001GaAs}.
The number of defects as predicted via the NRT model ($N_{\mathrm{defects}} = \frac{0.8E_{PKA}}{2E_d}$) is evaluated with $E_d=13$ eV.
A fit to the number of EPH-MD defects at energies up to 5 keV is shown in grey, which has an effective $E_d$ of 17.1 eV.
}
\label{fNdefects}
\end{figure}

We ran additional EPH-MD calculations on 0.5, 2, 5 and 20 keV PKAs. The evolution of the number of defects is shown in Figure \ref{fNdefects}.
A clear linear relationship is observed for the number of interstitials and antisites.
We also included the number of defects expected from the NRT formula: 
$N_{\mathrm{defects}} = \frac{0.8E_{PKA}}{2*E_d}$ 
where $E_d$ is the threshold displacement energy. $E_d$ is taken as 13 eV, the average of the reported values for Ga and As \cite{chen2017}.
As one can see, the number of defects, as obtained with the cut-off method is considerably higher than what predicted via the NRT model, but has the same linearity. 

The number of defects at 1 keV and 10 keV increases with approximately a factor of 9-10, in agreement to what is observed in MD simulations by Gao \textit{et al.} \cite{gao2017displacement} and Nordlund \textit{et al.} \cite{Nordlund2001GaAs}.
However, contrary to our work, in the work by Gao \textit{et al} \cite{gao2017displacement}, the number of defects was much higher than what predicted by the NRT model as shown in Figure \ref{fNdefects}.
The authors associated this results with the direct-impact amorphization that occurs in GaAs, which prevents the fast annihilation of interstitials with vacancies during the relaxation phase. The defect production efficiency, defined as the ratio of $N_F$ to $N_{NRT}$, ranged from 3.0 to 5.0 and increased with increasing PKA energy, rather than decreasing with increasing PKA energy as revealed in SiC and metals. This suggested that the defect production efficiency in GaAs is higher than those found in metals or alloys because of the absence of a thermal spike in GaAs.
When determined via the WS definition, our number of defects is slightly smaller than predicted via the NRT model. In both cases, there is a high linearity in the number of defects \textit{vs.}\ PKA energy.
The number of defects reported in Nordlund \textit{et al.} \cite{Nordlund2001GaAs} for GaAs is higher than ours but much closer to our reported values than those from Gao \textit{et al} \cite{gao2017displacement}.
As in their case, the number of antisites is an order of magnitude lower than the number of standard defects (vacancies and interstitials).
\textcolor{black}{We note that great care should be taken to the comparison of absolute defects numbers as there are still large uncertainties also indicated by Nordlund \textit{et al.} \cite{Nordlund2001GaAs} not only due to different electronic stopping models but also due to the use of different force field potentials and defect detection algorithms.
}

As can be seen from the linear fit to the lower energy defect numbers given in Figure \ref{fNdefects}, at higher energies, the agreement with the linear NRT model breaks down to some extent. This can be caused by the larger amount of energy lost by electronic stopping and also because of non-linear effects due to the start of overlapping defect cascades.

\section{Conclusions}

In this work, we have shown that it is possible to get proper \textit{ab-initio} electronic stopping behaviour of PKAs in GaAs via RT-TDDFT calculations. 
It has to be noted that while RT-TDDFT simulations are becoming more widely used, simulations of heavy ions are anything but routine.
We devised a more robust procedure as put forward by Caro \textit{et al.} \cite{Caro2019} to fit the EPH model to the ESP determined via RT-TDDFT. A different two-tier method was used that immediately prevents unbounded stopping and that relies on two global optimization steps to reduce the problem of local optima.
Although theoretically based on metals, we have shown that the EPH model provides a good description of the electronic stopping in the semiconductor material GaAs. Moreover, we have demonstrated the application of the EPH model for multiple elements materials.

We conclude that the inclusion of the electronic stopping and electron-phonon coupling influences the number of defects, and is thus crucial to correctly model the radiation damage.
The temperature dissociation in the TTM-MD was proven to be much faster, while the EPH model dissipated the heat spike much more slowly.

In comparison with the commonly used NRT model, the cut-off method gives a higher number of defects, while the Wigner-Seitz method gives a lower one.
Our work shows that the number of defects obtained with the EPH model differs from the results obtained with other models and within the Wigner-Seitz definition of defects, this number is "attenuated".
Although our work suggests that the realistic, non-adiabatic cascades, would not strongly affect the linear dependence of $N_{non-adiab}$ \textit{vs}.\ $E_{PKA}$, deviations from linearity start to increase beyond PKA energies of 10 keV.
Such non linear effects could be caused by the larger amount of energy lost by electronic stopping and also because of non-linear effects due to the start of overlapping defect cascades. Non linearity effects observed in experiments could also be induced by intrinsic defects in the target, an aspect that should be further investigated.
It has to be stressed that an overestimation of the number of defects at the nanometer scale can affect the lifetime estimation of the functions that the material serves to. 

\section{acknowledgements}

This work received funding from the Research Executive Agency under the EU's Horizon 2020 Research and Innovation program ESC2RAD (grant ID 776410). 
We acknowledge PRACE for awarding us access to Joliot-Curie at GENCI@CEA, France (PRACE Project: 2019215186).
The author thankfully acknowledges the computer resources at MareNostrum
and the technical support provided by Barcelona Supercomputing Center
(RES-FI-2020-1-0012, RES-FI-2020-2-0021, RES-FI-2020-3-0014,
RES-FI-2021-1-0035).

Jorge Kohanoff was supported by the Beatriz Galindo Program (BEAGAL18/00130) from the Ministerio de Educaci\'on y
Formaci\'on Profesional of Spain, and by the Comunidad de Madrid through the Convenio Plurianual with 
Universidad Polit\'ecnica de Madrid in its line of action Apoyo a la realizaci\'on de proyectos de I+D para
investigadores Beatriz Galindo, within the framework of V PRICIT (V Plan Regional de Investigaci\'on 
Cient\'ifica e Innovaci\'on Tecnol\'ogica).

We thank Alfredo Correa from the Lawrence Livermore National Laboratory for fruitful discussions.
Funding from Spanish MICIN is also acknowledged through grant PID2019-107338RB-C61/AEI/DOI:10.13039/501100011033 as well as a Mar\'{\i}a de Maeztu award to Nanogune, Grant CEX2020-001038-M funded by MCIN/AEI/10.13039/501100011033.

For Fabiana Da Pieve, the information and views set out in this article are those of the author and do not necessarily reflect the official opinion of the ERCEA.

\bibliography{biblio}

\end{document}